\documentclass[12pt,letterpaper]{article}
\usepackage{amsmath}
\usepackage{amsfonts}
\usepackage{amssymb}
\usepackage[margin=1in]{geometry}
\usepackage{graphicx}
\usepackage{float}
\usepackage{mathrsfs}
\usepackage{enumitem}
\usepackage{bm}
\usepackage{algorithm2e}
\usepackage{fancyhdr}
\usepackage{url}
\usepackage{subcaption}
\usepackage{mwe}
\usepackage{titling}

\begin{document}
%create title
\title{A Variational Inference Algorithm for BKMR in the Cross-Sectional Setting}
\author{Raphael Small\thanks{The authors gratefully acknowledge National Institutes of Health grants ES000002, ES028800, ES028811, and ES026555.} \and Brent A. Coull}
\date{Department of Biostatistics\\Harvard T.H. Chan School of Public Health\\[2ex]
	\today
}
\maketitle

%create abstract
\begin{abstract}
	\setlength{\parskip}{\baselineskip}
	\setlength{\parindent}{0pt}
	\noindent
	The identification of pollutant effects is an important task in environmental health. Bayesian kernel machine regression (BKMR) is a standard tool for inference of individual-level pollutant health-effects, and we present a mean field Variational Inference (VI) algorithm for quick inference when only a single response per individual is recorded. Using simulation studies in the case of informative priors, we show that VI, although fast, produces anti-conservative credible intervals of covariate effects and conservative credible intervals for pollutant effects. To correct the coverage probabilities of covariate effects, we propose a simple Generalized Least Squares (GLS) approach that induces conservative credible intervals. We also explore using BKMR with flat priors and find that, while slower than the case with informative priors, this approach yields uncorrected credible intervals for covariate effects with coverage probabilities that are much closer to the nominal 95\% level. We further note that fitting BKMR by VI provides a remarkable improvement in speed over existing MCMC methods.
\end{abstract}
\setlength{\parskip}{\baselineskip}
\setlength{\parindent}{0pt}

\newpage
\section*{Introduction}
The identification of pollutant effects plays an important role in the analysis of environmental health data. Often, pollutant exposures are highly correlated and have non-linear effects. Furthermore, individuals may only have a single response captured in an observational study, making standard longitudinal methods of inference not applicable.

To solve these problems, Bobb et al. (2015) introduced a Bayesian Kernel Machine Regression (BKMR) model in which pollutant effects are modeled as random effects with a covariance matrix that is parameterized via a kernel. Bobb et al. also include variable selection features using ``slab-and-spike" priors and demonstrate the effectiveness of their model. Liu et al. (2017) then extended the BKMR model to identify ``windows of susceptibility" and the effect of multi-pollutant exposures over time.

Since standard MCMC methods for Bayesian inference can be slow, Liu et al. (2018) introduced a mean field Variational Inference (VI) algorithm for faster inference in analyzing time-varying pollutant exposures. In this paper, we develop a VI algorithm for inference in the original BMKR formulation of cross-sectional data. See Blei et al. (2018), Gelman et al. (Chapter 13.7), and Murphy (Chapter 21) for an overview of VI.

As Westling and McCormick (2017) point out, variational approximations and their estimators have different properties than typical likelihood-based frequentist estimators. VI-derived credible intervals are known to often be too tight (see Wang and Titterington 2005) and Westling and McCormick show how Variational Bayes can result in coverage probabilities well below the nominal 95\% level. To overcome this problem, they connect VI to M-estimators and derive a sandwich estimator and MLE-based correction. 

The Westling and McCormick setting includes both latent variables and parameters, as is the case in BKMR, yet the correction method involves calculating Hessian matrices with respect to the unknown parameters and variational distribution parameters of the latent variables. Since BKMR estimates the covariance matrix of pollutant effects over all subjects, calculating a Hessian with respect to the covariance matrix can be computationally infeasible. In order to provide quick inference while keeping the desirable Bayesian properties of BKMR and restoring nominal coverage probabilities, we propose using the results of VI to estimate the covariance matrix in Generalized Least Squares (GLS).

In our simulation studies, we find that Variational Bayes with informative priors (VI1) has coverage probabilities for pollutant effects that exceed the nominal 95\% level but that are decreasing in conservativeness. When applying VI with flat priors (VI2), coverage probabilities can be extremely anti-conservative but also display significant increasing conservativeness as sample size rises.

In terms of coverage probabilities for covariate effects, we find that VI1 is anti-conservative but that the coverage probabilities rise slightly with increasing sample size. We also find that VI2 has coverage probabilities that cluster around 95\% except for the intercept. In applying our GLS correction, the modified results for VI1 and VI2 (GLS1 and GLS2 respectively) yield coverage probabilities that respectively exceed 95\% and cluster around 95\%. The GLS2 credible interval for the intercept is again anti-conservative. Furthermore, we also find that Variational Inference provides an orders of magnitude improvement in speed over existing MCMC methods for fitting BKMR models.

In short, we present experimental results that show Variational Inference for BMKR can yield fast results while achieving credible intervals for pollutant effects that are conservative. We also show that VI yields anti-conservative coverage probabilities for covariate effects, but that a GLS correction can lead to remarkably improved coverage probabilities.

\section*{The BKMR Model and Variational Inference}
Following Bobb et al., we formulate health outcomes as a linear combination of covariate and pollutant effects
\begin{equation}
y_i\sim\mathcal{N}\left(h_i(\textbf{z}_i)+\textbf{x}_i\bm\beta,\sigma^2\right)
\end{equation}
where $ h(\cdot) $ is the pollutant effect as some function of environmental exposures $ \textbf{z}_i $. Liu et al. (2007) showed the connection between kernel machine regression and linear mixed models in which $ h(\cdot) $ is written in its ``dual form" so that the model is operationalized by
\begin{equation}
\textbf{y}\sim\mathcal{N}\left(\textbf{h}+\textbf{X}\bm\beta, \sigma^2\textbf{I}\right)
\end{equation}
\begin{equation}
\textbf{h}\sim\mathcal{N}\left(\textbf{0},\tau\textbf{K}\right)
\end{equation}
where $ \textbf{K} $ is the positive definite ``kernel". Note that the likelihood is traditionally expressed as $ y_i $ from the equivalent univariate normal distribution. We use the multivariate normal here because it often makes derivations simpler in practice.

Each element of the kernel is defined as $ \textbf{K}_{ij}=K\left(\textbf{z}_i,\textbf{z}_j\right) $ where $ K $ is the kernel function that maps ``close" $ \textbf{z}_i $ and $ \textbf{z}_j $ to low covariances. This means that similar profiles of environmental exposures are assumed to have similar pollutant effects through the low covariance. Popular kernels include the linear, quadratic, and Gaussian (radial basis function). Since kernel inference can be difficult (see Liu and Coull 2017), we restrict our model to the quadratic kernel which does not require parameter specification. To ensure that the resulting kernel is positive definite, we employ the $ \texttt{nearPD} $ function from the $ \texttt{Matrix} $ R package which is based on the Higham (2002) algorithm.

We complete the Bayesian specification of the model by imposing a Gaussian prior for $ \bm\beta $ and inverse-gamma priors for $ \sigma^2 $ and $ \tau $ under the scaled-inverse chi-squared parameterizations
\begin{equation}
\bm\beta\sim\mathcal{N}\left(\bm\mu,\bm\Sigma\right)
\end{equation}
\begin{equation}
\sigma^2\sim\text{Scale-Inv-}\chi^2\left(\nu_\sigma,\sigma_0^2\right)
\end{equation}
\begin{equation}
\tau\sim\text{Scale-Inv-}\chi^2\left(\nu_\tau,\tau_0\right)
\end{equation}

As previously discussed, VI has been proposed as a method for quick inference to overcome the problem of slow convergence with MCMC methods. The efficiency gains from VI are due to the fact that it casts a sampling problem into an optimization problem. Specifically, an approximate posterior (q-density) is fit to the true posterior by minimizing the KL divergence between the two. The objective function is therefore
\begin{equation}
\mathcal{L}=\text{KL}(q||p)=-\mathbb{E}_q\left[\ln\dfrac{p(\bm\theta|\textbf{y})}{q(\bm\theta)}\right]=\mathbb{E}_q\left[\ln q(\bm\theta)\right]-\mathbb{E}_q\left[p(\bm\theta|\textbf{y})\right]
\end{equation}
where we treat $ \bm\theta $ as an amalgamation over the parameters $ \bm\beta $, $ \tau $, and $ \sigma^2 $ and the latent variables $ \textbf{h} $. $ q(\bm\theta) $ therefore represents the joint approximation of the posterior, while $ p(\bm\theta|\textbf{y}) $ is the true posterior. In treating the parameters and latent variables ``equally", we employ an approach termed ``Variational Bayes EM" by Murphy (pg 750).

One of the most popular forms of VI is the mean field approach, in which we assume a factored approximation $ q(\bm\theta)=q(\bm\beta)q(\textbf{h})q(\sigma^2)q(\tau) $. In mean field, each separate q-density is updated iteratively, leading to an algorithm that is sometimes called Coordinate Ascent Variational Inference (see Blei 2018) because each step decreases the KL divergence.

Rather than show the algebra of deriving the mean field VI updates, we state the results in Algorithm 1 on the next page. The full derivations can be found in the appendix.

\newpage
\begin{algorithm}[H]
	\caption{Compute approximate posterior with informative priors.}
	
	\KwIn{$ n $, $ \textbf{X} $, \textbf{y}, $ \bm\mu $, $ \bm\Sigma $, $ \nu_\sigma $, $ \sigma_0^2 $, $ \nu_\tau $, $ \tau_0 $, $ \textbf{K} $}
	\textbf{Initialize} $ \bm\mu_{q(\bm\beta)} $, $ \bm\Sigma_{q(\bm\beta)} $, $ \bm\mu_{q(\textbf{h})} $, $ \bm\Sigma_{q(\textbf{h})} $\\
		$ \nu_{q(\sigma^2)}:=n+\nu_\sigma $ \\
	$ \nu_{q(\tau)}:=n+\nu_\tau $ \\
	\While{$ \mathcal{L} $ has not converged}
	{
		$ D_{\sigma^2}\leftarrow\text{tr}\left(\bm\Sigma_{q(\textbf{h})}+\textbf{X}\bm\Sigma_{q(\bm\beta)}\textbf{X}^T\right)+\left(\textbf{y}-\bm\mu_{q(\textbf{h})}-\textbf{X}\bm\mu_{q(\bm\beta)}\right)^T\left(\textbf{y}-\bm\mu_{q(\textbf{h})}-\textbf{X}\bm\mu_{q(\bm\beta)}\right) $\\
		$ \sigma_{0,q(\sigma^2)}^2\leftarrow\left(D_{\sigma^2}+\nu_\sigma\sigma_0^2\right)/\nu_{q(\sigma^2)} $\\
		$  $\\
		$ D_{\tau}\leftarrow\text{tr}\left(\textbf{K}^{-1}\bm\Sigma_{q(\textbf{h})}\right)+\bm\mu_{q(\textbf{h})}^T\textbf{K}^{-1}\bm\mu_{q(\textbf{h})} $\\
		$ \tau_{0,q(\tau)}\leftarrow\left(D_\tau+\nu_\tau\tau_0\right)/\nu_{q(\tau)} $\\
		$  $\\
		$ \bm\Sigma_{q(\textbf{h})}\leftarrow\left(\textbf{I}/\sigma_{0,q(\sigma^2)}^2+\textbf{K}^{-1}/\tau_{0,q(\tau)}\right)^{-1} $\\
		$ \bm\mu_{q(\textbf{h})}\leftarrow\bm\Sigma_{q(\textbf{h})}\left(\textbf{y}-\textbf{X}\bm\mu_{q(\bm\beta)}\right)/\sigma_{0,q(\sigma^2)}^2 $\\
		$  $\\
		$ \bm\Sigma_{q(\bm\beta)}\leftarrow\left(\textbf{X}^T\textbf{X}/\sigma_{0,q(\sigma^2)}^2+\bm\Sigma^{-1}\right)^{-1} $\\
		$ \bm\mu_{q(\bm\beta)}\leftarrow\bm\Sigma_{q(\bm\beta)}\left(\textbf{X}^T\left(\textbf{y}-\bm\mu_{q(\textbf{h})}\right)/\sigma_{0,q(\sigma^2)}^2+\bm\Sigma^{-1}\bm\mu^T\right) $\\
		$  $\\
		Update the KL divergence $ \mathcal{L} $
	}
\end{algorithm}

\begin{algorithm}[H]
	\caption{Compute approximate posterior with flat priors.}
	
	\KwIn{$ n $, $ \textbf{X} $, \textbf{y}, $ \textbf{K} $}
	\textbf{Initialize} $ \bm\mu_{q(\bm\beta)} $, $ \bm\Sigma_{q(\bm\beta)} $, $ \bm\mu_{q(\textbf{h})} $, $ \bm\Sigma_{q(\textbf{h})} $\\
	$ \nu_{q(\sigma^2)}:=n-2 $ \\
	$ \nu_{q(\tau)}:=n-2 $ \\
	\While{$ \mathcal{L} $ has not converged}
	{
		$ D_{\sigma^2}\leftarrow\text{tr}\left(\bm\Sigma_{q(\textbf{h})}+\textbf{X}\bm\Sigma_{q(\bm\beta)}\textbf{X}^T\right)+\left(\textbf{y}-\bm\mu_{q(\textbf{h})}-\textbf{X}\bm\mu_{q(\bm\beta)}\right)^T\left(\textbf{y}-\bm\mu_{q(\textbf{h})}-\textbf{X}\bm\mu_{q(\bm\beta)}\right) $\\
		$ \sigma_{0,q(\sigma^2)}^2\leftarrow D_{\sigma^2}/\nu_{q(\sigma^2)} $\\
		$  $\\
		$ D_{\tau}\leftarrow\text{tr}\left(\textbf{K}^{-1}\bm\Sigma_{q(\textbf{h})}\right)+\bm\mu_{q(\textbf{h})}^T\textbf{K}^{-1}\bm\mu_{q(\textbf{h})} $\\
		$ \tau_{0,q(\tau)}\leftarrow D_\tau/\nu_{q(\tau)} $\\
		$  $\\
		$ \bm\Sigma_{q(\textbf{h})}\leftarrow\left(\textbf{I}/\sigma_{0,q(\sigma^2)}^2+\textbf{K}^{-1}/\tau_{0,q(\tau)}\right)^{-1} $\\
		$ \bm\mu_{q(\textbf{h})}\leftarrow\bm\Sigma_{q(\textbf{h})}\left(\textbf{y}-\textbf{X}\bm\mu_{q(\bm\beta)}\right)/\sigma_{0,q(\sigma^2)}^2 $\\
		$  $\\
		$ \bm\Sigma_{q(\bm\beta)}\leftarrow\left(\textbf{X}^T\textbf{X}\right)^{-1}\sigma_{0,q(\sigma^2)}^2 $\\
		$ \bm\mu_{q(\bm\beta)}\leftarrow\bm\Sigma_{q(\bm\beta)}\left(\textbf{X}^T\left(\textbf{y}-\bm\mu_{q(\textbf{h})}\right)/\sigma_{0,q(\sigma^2)}^2\right) $\\
		$  $\\
		Update the KL divergence $ \mathcal{L} $
	}
\end{algorithm}

\section*{Simulation Studies and GLS Results}
We obtain cross-sectional data from the CDC's National Health and Nutrition Examination Survey (NHANES 2015-2016) and import the data into R using the \texttt{foreign} package. Using the observed empirical standard deviation, we simulate the systolic blood pressure of a hypothetical population subject to covariate and pollutant effects. We select Selenium (Se), Cadmium (Cd), Lead (Pb), and Mercury (Hg) as pollutants and set the pollutant effect as $ h_i=\text{Se}_i/100+\text{Cd}_i\text{Pb}_i+1/\text{Hg}_i-3 $. As illustrated in Figure 1, the distribution of effects is extremely non-Gaussian. Using the simulated data, we sample datasets of size $ n=100,200,300,400,500 $ observations without replacement. For each $ n $, we perform 1,000 such re-samplings and pass each dataset through the VI algorithm to examine the characteristics of the estimators.

We use a simple prior elicitation strategy. For each sampled dataset we regress the sampled $ \textbf{y} $ on $ \textbf{X} $ and take $ \bm\mu $ as the set of estimated coefficients and take $ \bm\Sigma $ as the resulting variance-covariance matrix. We set $ \nu_\sigma $ as the residual degrees of freedom and $ \sigma_0^2 $ to the regression estimate of the variance. Lastly, we impose a vague prior on $ \tau $ with $ \tau_0=1 $ and $ \nu_\tau=10 $. Restricted Maximum Likelihood can be used to estimate $ \sigma $ and $ \tau $ (as well as the Gaussian tuning parameter $ \rho $ of $ \textbf{K} $ if need be), but such methods become expensive with even moderate $ n $ (see Liu et al. 2007 and Liu and Coull 2017) which is why we resort to the OLS elicitation strategy here.

It is clear that $ \bm\mu_{q(\bm\beta)} $ and $ \bm\mu_{q(\textbf{h})} $ are natural point estimates for the true $ \bm\beta $ and $ \textbf{h} $. Similarly, $ \text{diag}\left(\bm\Sigma_{q(\bm\beta)}\right)^{1/2} $ and $ \text{diag}\left(\bm\Sigma_{q(\textbf{h})}\right)^{1/2} $ are our standard errors when building Wald-type credible intervals. For obtaining point estimates of $ \sigma^2 $, we use the mode of the scaled-inverse chi-squared distribution, which occurs at $ \nu_{q(\sigma^2)}\sigma_{0,q(\sigma^2)}/\left(\nu_{q(\sigma^2)}+2\right) $. The point estimate of $ \sigma^2 $ is therefore a MAP estimate in the sense of the approximate variational posterior.

The resulting coverage probabilities of the VI credible intervals for covariate effects are presented in Table 1 and displayed in Figure 2. Applying mean field VI for BKMR (VI1) yields coverage probabilities that are far below the 95\% nominal level. If we are unwilling (or unable) to impose priors, we derive a simpler VI algorithm (VI2 in Algorithm 2) using flat priors. Although VI2 takes longer to converge, it does achieve coverage probabilities that can be quite close to 95\% with the exception of the intercept $ \beta_0 $. Although the coverage probabilities of VI1 tend to increase as $ n $ rises, they remain relatively flat for VI2. It should be noted that for each time we run VI1 and VI2, we limit the number of iterations to 500 and use a convergence criterion of $ 10^{-2} $ when tracking $ \mathcal{L} $. The results for VI2 may therefore be subject to insufficient convergence.

To maintain the benefits of a Bayesian approach while recovering nominal coverage probabilities, we apply the VI1 results to GLS. Using the BKMR likelihood as denoted by Equations 2 and 3, we write the health outcomes as
\begin{equation}
\textbf{y}=\textbf{h}+\textbf{X}\bm\beta+\bm\epsilon
\end{equation}
After performing VI, we have a posterior estimate of the distribution of $ \textbf{h} $ and a MAP estimate of $ \sigma^2 $. Under $ q(\bm\theta) $, holding $\bm\beta$ fixed at its true unknown value, the ``variational" distribution of $ \textbf{y} $ becomes
\begin{equation}
\textbf{y}\overset{q}{\sim}\mathcal{N}\left(\bm\mu_{q(\textbf{h})}+\textbf{X}\bm\beta,\bm\Sigma_{q(\textbf{h})}+\hat{\sigma}^2\textbf{I}\right)
\end{equation}
If we let $ \bm\Sigma_{q(\textbf{y})}=\bm\Sigma_{q(\textbf{h})}+\hat{\sigma}^2\textbf{I} $ and treat Equation 9 as a standard likelihood, we obtain the MLE result
\begin{equation}
\hat{\bm\beta}_{\text{GLS}}=\left(\textbf{X}^T\bm\Sigma_{q(\textbf{y})}^{-1}\textbf{X}\right)^{-1}\textbf{X}^T\bm\Sigma_{q(\textbf{y})}^{-1}\left(\textbf{y}-\bm\mu_{q(\textbf{h})}\right)
\end{equation}
which is the GLS result if $ \textbf{z}=\textbf{y}-\bm\mu_{q(\textbf{h})} $ is regressed on $ \textbf{X} $ with known covariance matrix $ \bm\Sigma_{q(\textbf{y})} $. If we assume that $ \bm\Sigma_{q(\textbf{y})} $ is the true covariance of $ \textbf{y} $, and holding $ \bm\mu_{q(\textbf{h})} $ fixed, then credible intervals can be obtained via
\begin{equation}
\hat{\bm\beta}_{\text{GLS}}\pm 1.96\times \text{diag}\left(\left[\textbf{X}^T\bm\Sigma_{q(\textbf{y})}^{-1}\textbf{X}\right]^{-1}\right)^{1/2}
\end{equation}
When applying this modification to the VI1 and VI2 results (GLS1 and GLS2 respectively), we see a significant improvement of coverage probabilities for GLS1 relative to VI1. The GLS1 credible intervals are conservative and slightly increase in conservativeness with $ n $ while the GLS2 CI's still hover at just below 95\%. Again, the intercept coverage probabilities are lower in GLS2, although the intercept coverage probability is higher in GLS2 vs VI2.

As for the coverage probabilities of the pollutant effects, we display the overall coverage probabilities (aggregated across individuals) in Table 2. The VI1 credible intervals are all conservative, but the degree of conservativeness declines slightly with $ n $. In contrast, the CI's for VI2 can be extremely anti-conservative but display a strong pattern of the coverage probabilities increasing. A possible explanation for this pattern may be that, without informative priors, VI2 must learn about both $ \bm\beta $ and $ \textbf{h} $ simultaneously which requires adequate sample size. In contrast, the OLS prior for $ \bm\beta $ in VI1 is likely a decent estimate for $ \bm\beta $, so there is more information to learn about pollutant effects.

In terms of the characteristics of $ \hat{\sigma}^2 $ in Table 3, we see that VI1 performs poorly both in terms of MSE and bias. The poor performance of VI1 in this regard could have been expected because the prior elicitation strategy for $ \sigma^2 $ ignores potential pollutant effect. This illustrates the importance of selecting an appropriate prior for $ \sigma^2 $. Decreasing $ \nu_\sigma $ or estimation via REML represent two possible solutions. The unexpected way in which the MSE for VI2 increases for $ n=400 $ and $ n=500 $ also illustrates the stabilizing effect of prior selection. Note that the degree of bias remains constant for VI1 while MSE is uniformly decreasing in sample size.

We also compare the computational performance of our VI-GLS approach to that of the typical MCMC approach. As a baseline, we use run times on NHANES data noted by Coull (2018) for BKMR fit with MCMC via Bobb's \texttt{bkmr} package in R (see Table 4 for further detail). Although we did not incorporate variable selection into our model (which would likely increase run time), our model is orders of magnitude faster, taking seconds to fit. All calculations were performed on a Windows 10 machine with an Intel i7-7500U processor (2.70GHz).

The baseline results in Table 4 were obtained from a dataset of $ n=1003 $. To match this sample size, we sampled 100 datasets of $ n=1003 $ from our NHANES data and recorded the run time for each. While \texttt{bmkr} accelerates performance with the introduction of $ k $ Gaussian predictive process knots, the fastest run time is approximately 54  minutes. In contrast, our VI model took an average of 21.22 seconds, with a minimum of 19.75 seconds and a maximum of 26.81 seconds. We include eleven covariates in addition to an intercept term and four pollutants. We use a stringent convergence criterion of $ 10^{-6} $ on $ \mathcal{L} $ and impose a mandatory ``burn-in" of 10 iterations. Each sampled dataset required eleven iterations before reaching convergence. For comparison, forcing the algorithm to perform 100 and 500 iterations required approximately 3.3 and 16.7 minutes respectively.

\section*{Conclusion}
Using simulation studies, we have shown that Variational Inference can result in anti-conservative credible intervals with respect to covariate effects. When applying VI to BKMR with flat priors, we find that the covariate effect coverage probabilities, excluding the intercept, are much closer to the nominal 95\% level, illustrating the importance of prior selection. Although we did not explore it in this paper, modifying the Restricted Maximum Likelihood approach of Liu et al. (2007) and Liu and Coull (2017) for improved speed may yield useful starting points for choosing priors.

One important result of this paper is that VI, although very fast, can yield credible intervals that are too tight, so variational approximations to the posterior are not a panacea. Despite the limitations of VI, the coverage probabilities in our simulation studies had a minimum of about 80\%, which may be an acceptable price to allow the researcher to rapidly explore many models. 

In addition, we provide experimental evidence that shows how VI can be combined with Generalized Least Squares to provide coverage probabilities that can be very close to the nominal 95\% level (in the case of flat priors) or exceed it (in the case of informative priors). The VI-GLS approach therefore maintains the speed of Variational Inference while yielding trustworthy, i.e. conservative, credible intervals for covariate effects. Further understanding the properties of variational estimators for BKMR, perhaps with larger and more diverse simulation studies, should be an area of future research.

In our simulation studies, we find that the coverage probabilities for pollutant health-effects exceeds 95\% even for $ n=100 $ in the case of informative priors. Without informative priors, sufficient sample size is required for valid inferences. This further validates the usefulness of BKMR (with priors) in identifying pollutant health effects and the application of VI to the problem. Furthermore,  we have demonstrated the effectiveness of our VI algorithm for identifying multi-pollutant health-effects in the cross-sectional setting. The improvement in computing time over traditional MCMC methods also makes VI an appealing method for fitting BKMR models.

We also find that VI with informative priors is a superior approach to VI with flat priors, even though the non-intercept coverage probabilities of VI2 exceed those of VI1. We make this judgment based on four reasons. First, our GLS correction yields conservative credible intervals under informative priors. In contrast, the GLS correction for VI2 does not lead to significant improvements for coverage probabilities, and the CI for the intercept remains anti-conservative. Second, our informative priors yield credible intervals for pollutant effects that are conservative even with small sample sizes, in contrast with the case of flat priors. Thirdly, flat priors can result in unpredictable results in the estimation of the residual variance $ \sigma^2 $. Fourthly, our algorithm with informative priors converges much faster.

In conclusion, we have introduced a mean field Variational Inference algorithm for inference in Bayesian Kernel Machine Regression to identify pollutant effects in the cross-sectional setting. We demonstrate that VI yields conservative credible intervals for pollutant effects but anti-conservative intervals for covariate effects. Practitioners should therefore be aware that the speed of VI comes at a cost. To overcome this limitation, we show that VI results under a simple prior elicitation strategy can be combined with Generalized Least Squares to obtain conservative credible intervals for covariate effects. As we have shown, Variational Inference is a powerful tool for identifying pollutant effects, yet further study is warranted to better understand its properties under Bayesian Kernel Machine Regression.

\newpage
\section*{Tables}

\begin{table}[H]
	\centering
	\begin{tabular}{|lcccccc|}
		\hline
		& $\beta_0$ & $\beta_1$ & $\beta_2$ & $\beta_3$ & $\beta_4$ & $\beta_5$ \\\hline\hline
		$n=100$ &           &           &           &           &           &           \\
		VI1     & 0.819     & 0.814     & 0.834     & 0.803     & 0.819     & 0.797     \\
		VI2     & 0.859     & 0.938     & 0.945     & 0.943     & 0.937     & 0.945     \\
		GLS1    & 0.978     & 0.973     & 0.979     & 0.972     & 0.979     & 0.970     \\
		GLS2    & 0.869     & 0.939     & 0.951     & 0.944     & 0.937     & 0.944     \\
		&           &           &           &           &           &           \\
		$n=200$ &           &           &           &           &           &           \\
		VI1     & 0.818     & 0.850     & 0.819     & 0.831     & 0.793     & 0.831     \\
		VI2     & 0.847     & 0.959     & 0.936     & 0.957     & 0.925     & 0.953     \\
		GLS1    & 0.982     & 0.986     & 0.976     & 0.988     & 0.976     & 0.975     \\
		GLS2    & 0.862     & 0.960     & 0.943     & 0.960     & 0.933     & 0.955     \\
		&           &           &           &           &           &           \\
		$n=300$ &           &           &           &           &           &           \\
		VI1     & 0.831     & 0.853     & 0.843     & 0.825     & 0.830     & 0.838     \\
		VI2     & 0.824     & 0.958     & 0.935     & 0.950     & 0.927     & 0.953     \\
		GLS1    & 0.977     & 0.987     & 0.981     & 0.980     & 0.984     & 0.981     \\
		GLS2    & 0.851     & 0.968     & 0.943     & 0.953     & 0.942     & 0.960     \\
		&           &           &           &           &           &           \\
		$n=400$ &           &           &           &           &           &           \\
		VI1     & 0.823     & 0.860     & 0.831     & 0.861     & 0.823     & 0.865     \\
		VI2     & 0.811     & 0.959     & 0.929     & 0.963     & 0.927     & 0.968     \\
		GLS1    & 0.979     & 0.985     & 0.975     & 0.984     & 0.977     & 0.984     \\
		GLS2    & 0.835     & 0.967     & 0.943     & 0.967     & 0.941     & 0.974     \\
		&           &           &           &           &           &           \\
		$n=500$ &           &           &           &           &           &           \\
		VI1     & 0.835     & 0.873     & 0.836     & 0.839     & 0.841     & 0.874     \\
		VI2     & 0.823     & 0.962     & 0.928     & 0.943     & 0.921     & 0.973     \\
		GLS1    & 0.985     & 0.993     & 0.985     & 0.981     & 0.989     & 0.990     \\
		GLS2    & 0.864     & 0.969     & 0.951     & 0.955     & 0.948     & 0.980     \\\hline
	\end{tabular}
	\caption{Coverage probabilities for covariate effects across both Variational Inference and Generalized Least Squares methods at different sample sizes.}
\end{table}

\begin{table}[H]
	\centering
	\begin{tabular}{|lcc|}
		\hline
		$n$ & VI1   & VI2   \\\hline\hline
		100 & 0.988 & 0.579 \\
		200 & 0.985 & 0.786 \\
		300 & 0.983 & 0.928 \\
		400 & 0.982 & 0.972 \\
		500 & 0.981 & 0.993 \\\hline
	\end{tabular}
	\caption{Aggregated coverage probabilities for pollutant effects for both VI algorithms.}
\end{table}

\begin{table}[H]
	\centering
\begin{tabular}{|lcccccc|}
	\hline
	& Mean   & SD    & 2.5\%-ile & Median & 97.5\%-ile & MSE     \\\hline\hline
	$n=100$ &        &       &           &        &            &         \\
	VI1     & 85.39  & 11.82 & 63.48     & 85.03  & 109.38     & 3795.56 \\
	VI2     & 101.94 & 14.33 & 75.38     & 101.34 & 130.96     & 2245.86 \\
	&        &       &           &        &            &         \\
	$n=200$ &        &       &           &        &            &         \\
	VI1     & 85.22  & 7.66  & 71.66     & 85.20  & 102.28     & 2978.40 \\
	VI2     & 100.56 & 9.31  & 83.60     & 100.65 & 120.02     & 934.03  \\
	&        &       &           &        &            &         \\
	$n=300$ &        &       &           &        &            &         \\
	VI1     & 85.41  & 6.01  & 73.60     & 85.37  & 97.13      & 2677.45 \\
	VI2     & 97.58  & 7.25  & 84.13     & 97.39  & 111.98     & 628.32  \\
	&        &       &           &        &            &         \\
	$n=400$ &        &       &           &        &            &         \\
	VI1     & 85.21  & 5.06  & 76.16     & 85.22  & 95.13      & 2626.77 \\
	VI2     & 94.19  & 6.38  & 82.84     & 94.08  & 107.57     & 800.30  \\
	&        &       &           &        &            &         \\
	$n=500$ &        &       &           &        &            &         \\
	VI1     & 85.19  & 4.40  & 77.07     & 85.01  & 93.77      & 2566.96 \\
	VI2     & 91.02  & 5.89  & 80.75     & 90.44  & 104.09     & 1239.83 \\\hline
\end{tabular}
	\caption{Analysis of bias for MAP estimates of $ \sigma^2 $. Note that bias as measured as $ 100\hat{\sigma}^2/\sigma^2 $ while MSE is reported on the actual error basis of $ \hat{\sigma}^2-\sigma^2 $.}
\end{table}

\begin{table}[H]
	\centering
	\begin{tabular}{|lcc|}
		\hline
		Model (Hours)     & Variable Selection & Hierarchical Variable Selection \\\hline\hline
		Full BKMR         & 7.0                & 5.6                             \\
		GPP ($k=100$)     & 1.4                & 1.3                             \\
		GPP ($k=50$)      & 0.9                & 0.9                             \\\hline\hline
		
		Process (Seconds) & Mean (SD)          & Range                           \\\hline\hline
		Prior Elictation  &   6.07 (0.17)                 &   5.91--7.45                              \\
		VI1 (Informative Priors)               &   21.22 (1.35)                &    19.75--26.81                            \\
		GLS CI Correction &         0.78 (0.06)          &      0.73--1.11                           \\\hline
	\end{tabular}
	\caption{A comparison of run times between BKMR fit using the \texttt{bkmr} package and our Variational Inference-Generalized Least Squares Approach. The baseline results are from another NHANES dataset with $n=1003$ (see Coull 2018). As our NHANES data had more observations, we sampled 100 datasets of size $n=1003$ and applied our model. Note the significant decrease in run time when using VI instead of MCMC. We included the creation of the kernel matrix $ \textbf{K} $ and its inversion in the prior elicitation process for the purposes of timing. Our algorithm terminated upon the convergence of $ \mathcal{L} $ using a convergence criterion of $ 10^{-6} $.}
\end{table}

\newpage
\section*{Figures}

\begin{figure}[H]
	\centering
	\includegraphics[width = 5 in]{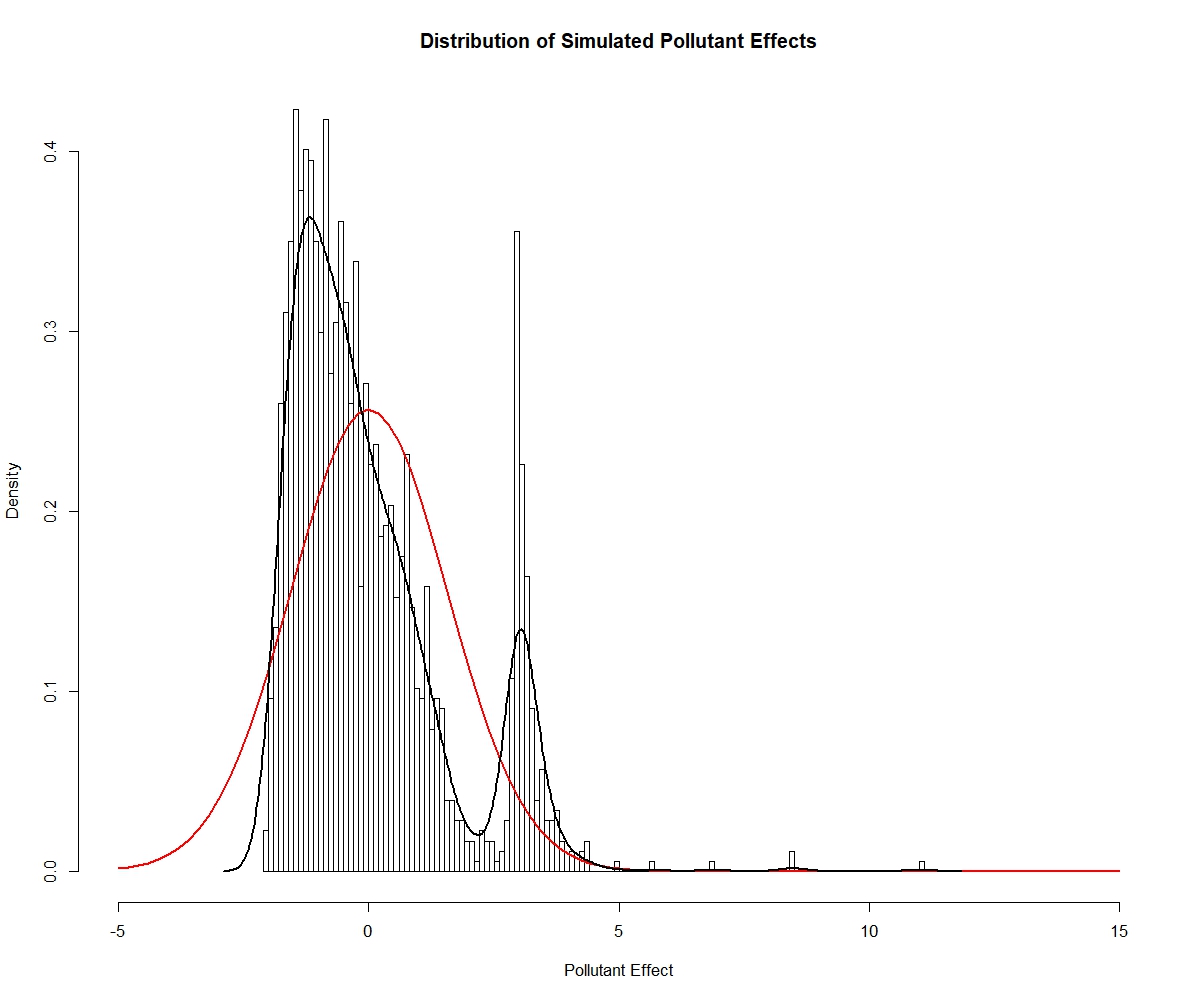}
	\caption{Note that the simulated population distribution of pollutant effects $ \textbf{h} $ is highly non-Gaussian. The black curve represents a Gaussian kernel density estimate. The red curve represents a the Gaussian density with the mean and standard deviation of $ \textbf{h} $.}
\end{figure}

\begin{figure*}
	\centering
	\begin{subfigure}[b]{0.475\textwidth}
		\centering
		\includegraphics[width=\textwidth]{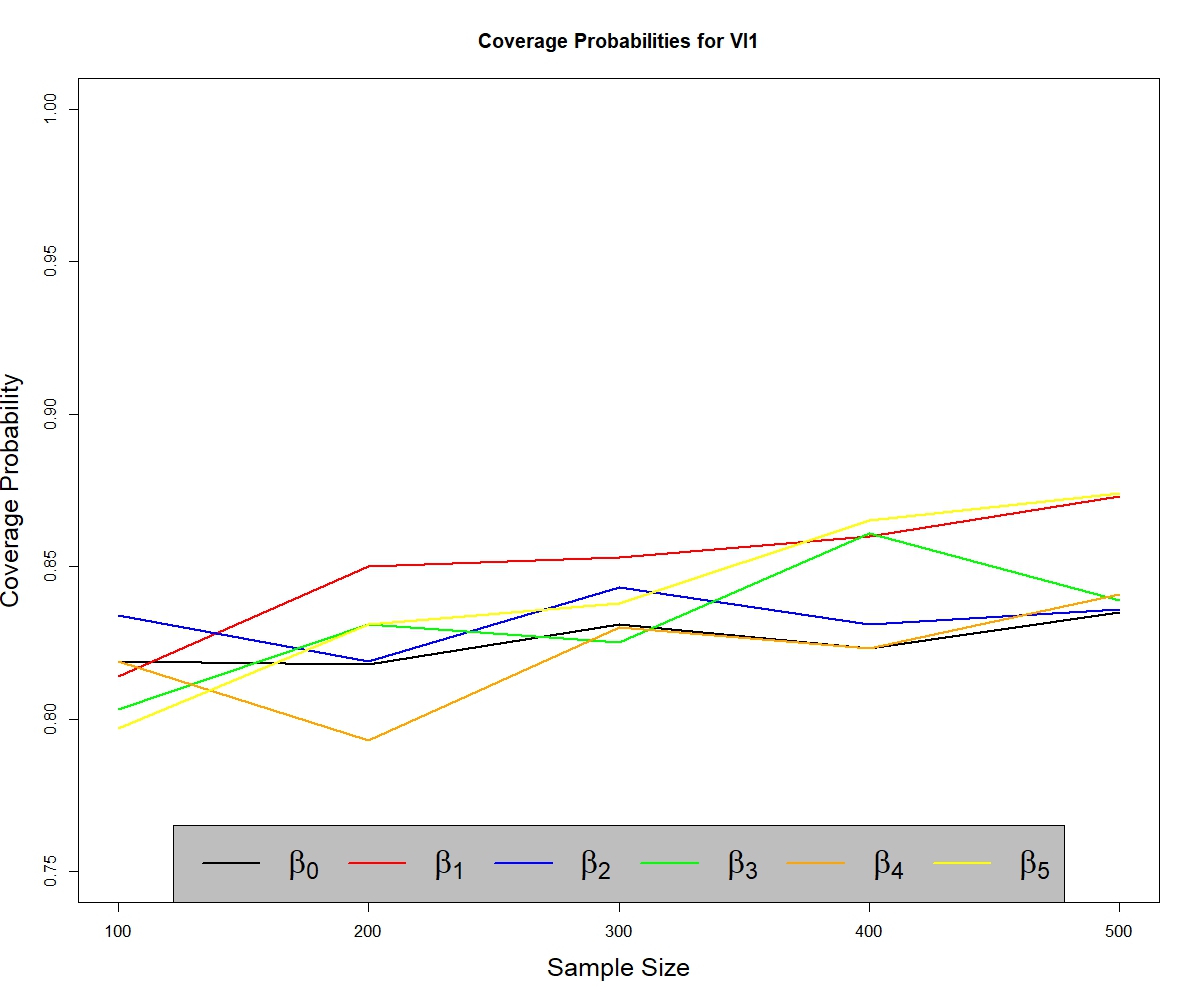}
	\end{subfigure}
	\hfill
	\begin{subfigure}[b]{0.475\textwidth}  
		\centering 
		\includegraphics[width=\textwidth]{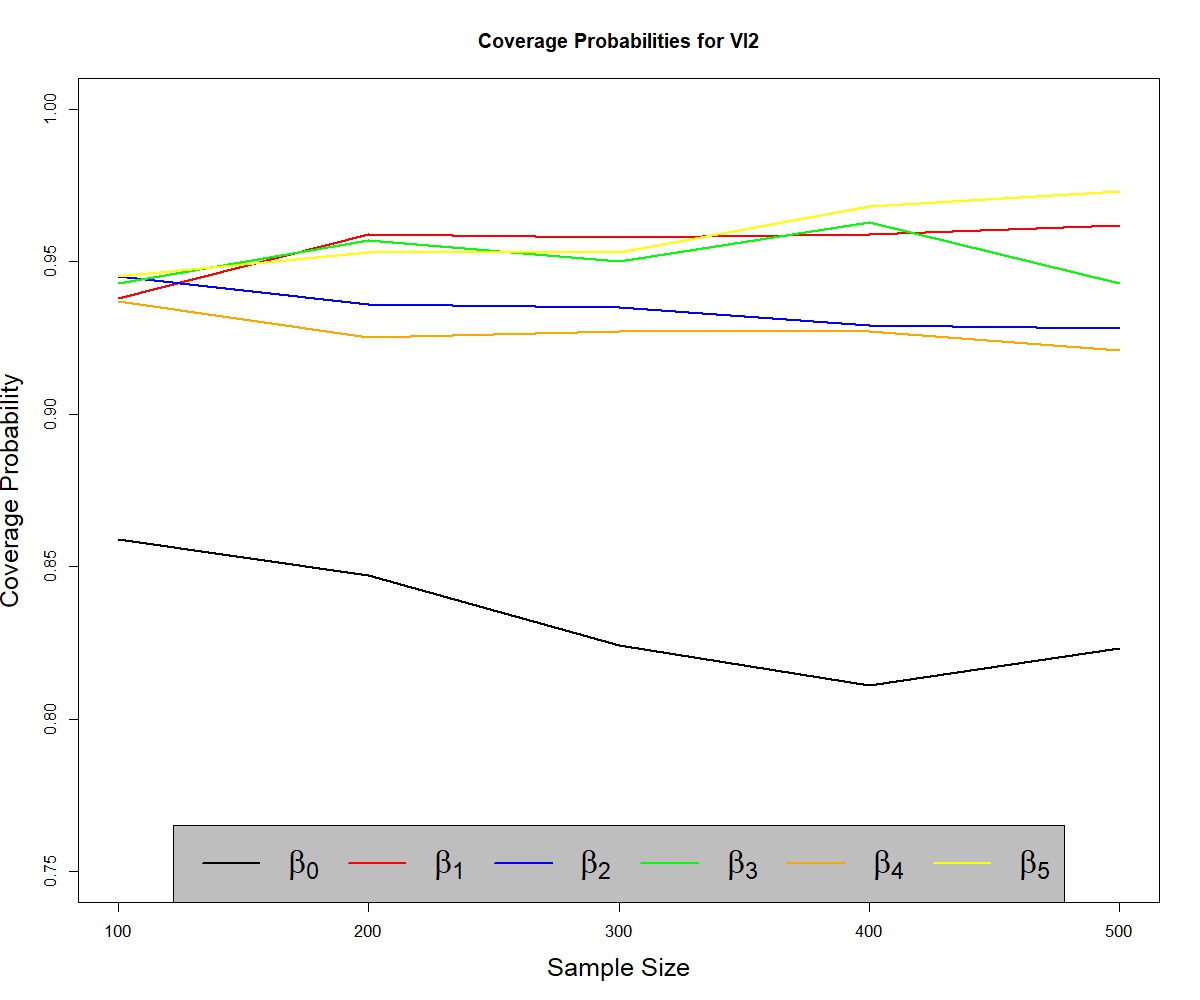}
	\end{subfigure}
	\vskip\baselineskip
	\begin{subfigure}[b]{0.475\textwidth}   
		\centering 
		\includegraphics[width=\textwidth]{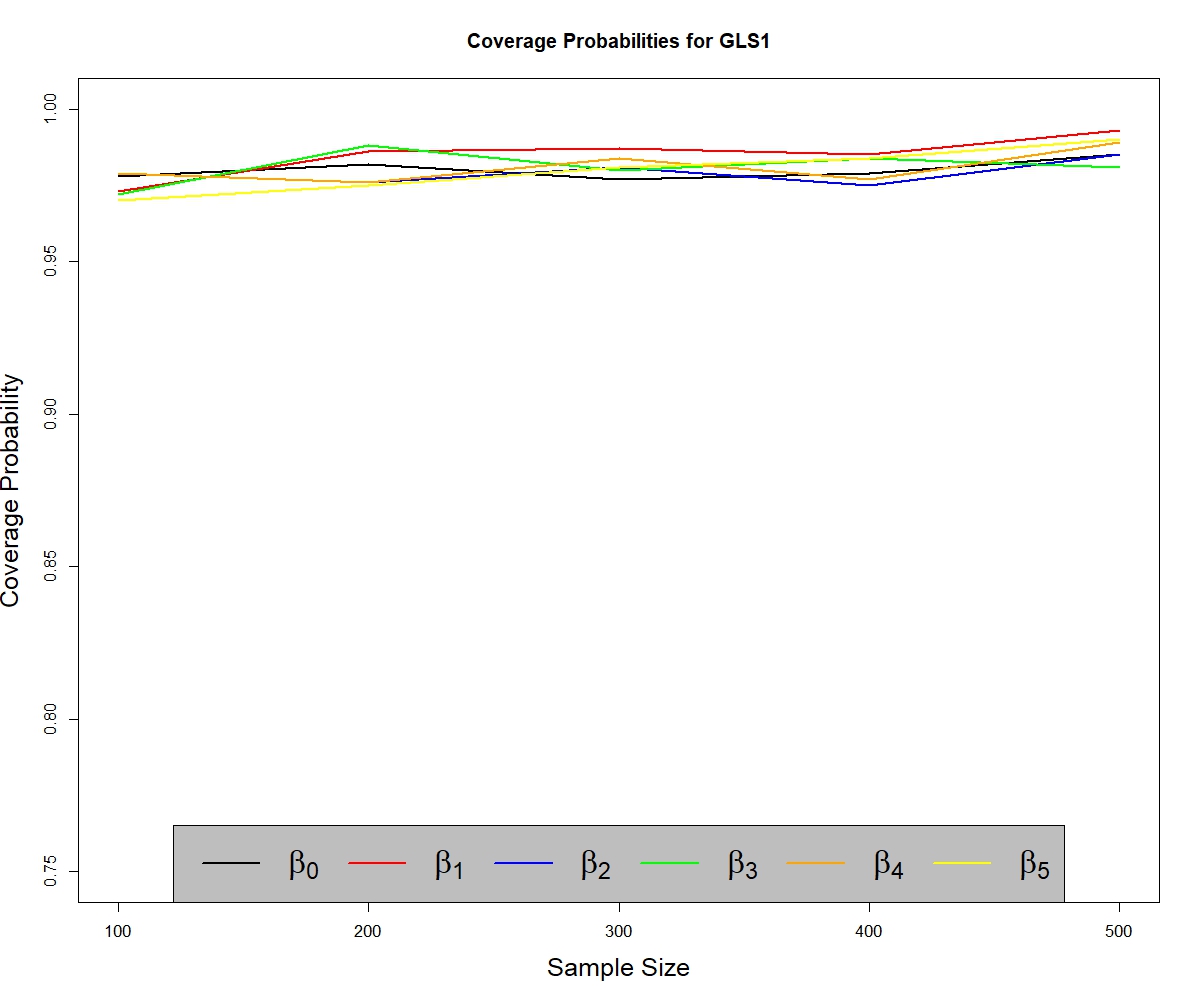}  
	\end{subfigure}
	\quad
	\begin{subfigure}[b]{0.475\textwidth}   
		\centering 
		\includegraphics[width=\textwidth]{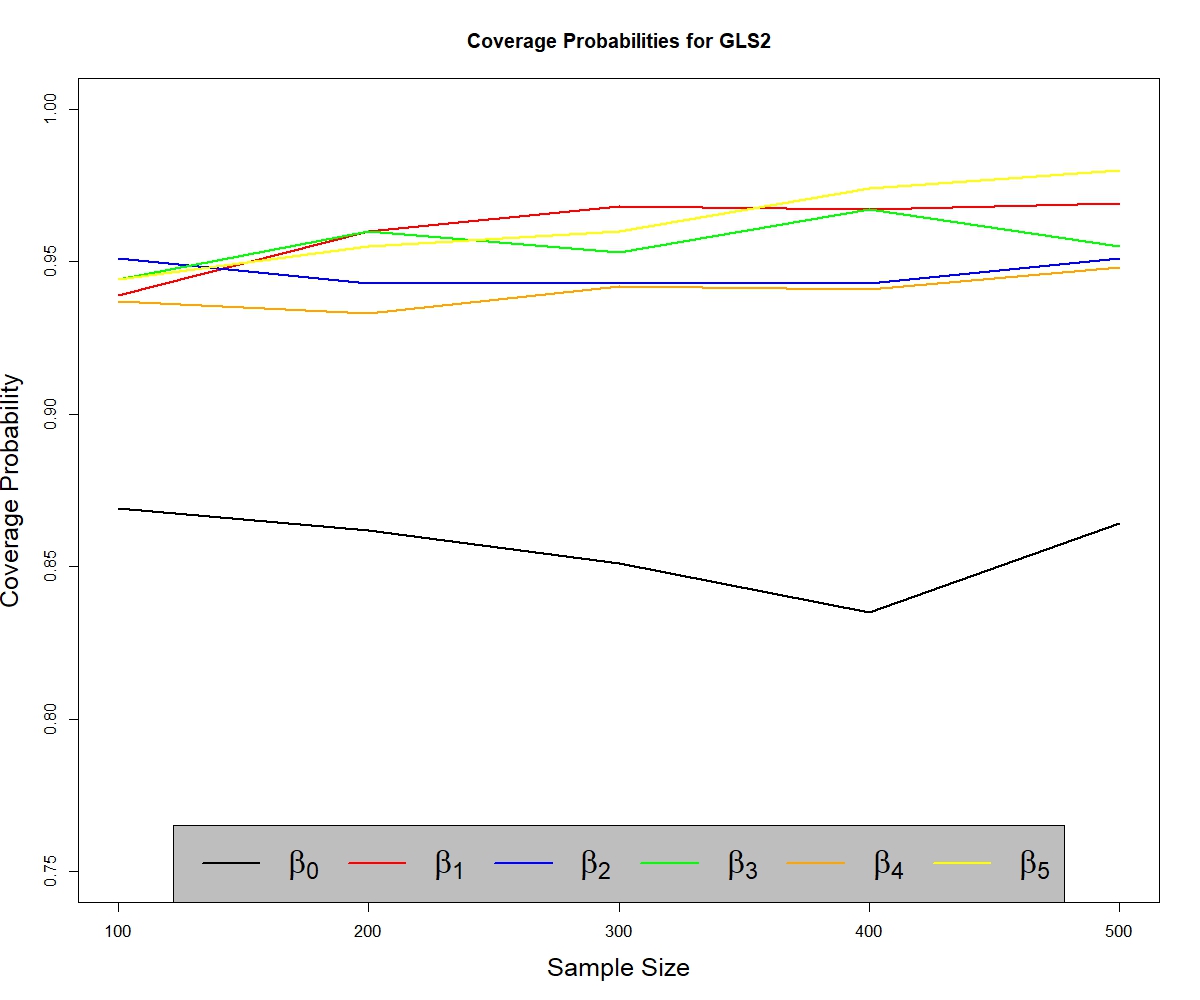}
	\end{subfigure}
	\caption{Coverage probabilities for covariate effects $ \bm\beta $ across both methods of inference.} 
\end{figure*}

\newpage
\section*{Appendix: VI Updates with Informative Priors}
\subsection*{VI Update for $ \bm\beta $}
Beginning with the MFVI updates for $ \bm\beta $, the conditional posterior is
\begin{align}
\pi\left(\bm\beta|\textbf{h},\sigma^2\right)\propto\exp\left[-\dfrac{1}{2}\left(\textbf{y}-\textbf{h}-\textbf{X}\bm\beta\right)^T\left(\textbf{y}-\textbf{h}-\textbf{X}\bm\beta\right)/\sigma^2\right]
\\\nonumber\times\exp\left[-\dfrac{1}{2}\left(\bm\beta-\bm\mu\right)^T\bm\Sigma^{-1}\left(\bm\beta-\bm\mu\right)\right]
\end{align}
The expectation of the log conditional posterior is therefore
\begin{align}
\mathbb{E}_{-q(\bm\beta)}\left[\ln\pi\left(\bm\beta|\textbf{h},\sigma^2\right)\right]=-\dfrac{1}{2}\mathbb{E}_{q(\textbf{h})}\left[\left(\textbf{y}-\textbf{h}-\textbf{X}\bm\beta\right)^T\left(\textbf{y}-\textbf{h}-\textbf{X}\bm\beta\right)\right]\times\mathbb{E}_{q(\sigma^2)}\left[1/\sigma^2\right]
\\\nonumber-\dfrac{1}{2}\left(\bm\beta-\bm\mu\right)^T\bm\Sigma^{-1}\left(\bm\beta-\bm\mu\right)+C
\\\nonumber=-\dfrac{1}{2}\left[\text{tr}\left(\mathbb{V}_{q(\textbf{h})}\left[\textbf{h}\right]\right)+\left(\textbf{y}-\mathbb{E}_{q(\textbf{h})}\left[\textbf{h}\right]-\textbf{X}\bm\beta\right)^T\left(\textbf{y}-\mathbb{E}_{q(\textbf{h})}\left[\textbf{h}\right]-\textbf{X}\bm\beta\right)\right]\times\mathbb{E}_{q(\sigma^2)}\left[1/\sigma^2\right]\\\nonumber-\dfrac{1}{2}\left(\bm\beta-\bm\mu\right)^T\bm\Sigma^{-1}\left(\bm\beta-\bm\mu\right)+C
\\\nonumber=-\dfrac{1}{2}\left(\textbf{y}-\mathbb{E}_{q(\textbf{h})}\left[\textbf{h}\right]-\textbf{X}\bm\beta\right)^T\left(\textbf{y}-\mathbb{E}_{q(\textbf{h})}\left[\textbf{h}\right]-\textbf{X}\bm\beta\right)\times\mathbb{E}_{q(\sigma^2)}\left[1/\sigma^2\right]-\dfrac{1}{2}\left(\bm\beta-\bm\mu\right)^T\bm\Sigma^{-1}\left(\bm\beta-\bm\mu\right)+C
\end{align}

Introducing some new notation, $ \textbf{d}=\textbf{y}-\mathbb{E}_{q(\textbf{h})}\left[\textbf{h}\right] $ and $ \Sigma_0^{-1}=\textbf{I}_n\times\mathbb{E}_{q(\sigma^2)}\left[1/\sigma^2\right] $, the entire expression can be simplified into a single quadratic form
\begin{equation}
\mathbb{E}_{-q(\bm\beta)}\left[\ln\pi\left(\bm\beta|\textbf{h},\sigma^2\right)\right]=-\dfrac{1}{2}\left[\bm\beta^T\left(\textbf{X}^T\bm\Sigma_0^{-1}\textbf{X}+\bm\Sigma^{-1}\right)\bm\beta-2\left(\textbf{d}^T\bm\Sigma_0^{-1}\textbf{X}+\bm\mu^T\bm\Sigma^{-1}\right)\bm\beta+C\right]
\end{equation}

Letting $ \textbf{A}=\textbf{X}^T\bm\Sigma_0^{-1}\textbf{X}+\bm\Sigma^{-1} $ and $ \textbf{B}=-2\left(\textbf{d}^T\bm\Sigma_0^{-1}\textbf{X}+\bm\mu^T\bm\Sigma^{-1}\right)^T $, the square can be completed as
\begin{equation}
\mathbb{E}_{-q(\bm\beta)}\left[\ln\pi\left(\bm\beta|\textbf{h},\sigma^2\right)\right]=-\dfrac{1}{2}\left(\bm\beta+\dfrac{1}{2}\textbf{A}^{-1}\textbf{B}\right)^T\textbf{A}\left(\bm\beta+\dfrac{1}{2}\textbf{A}^{-1}\textbf{B}\right)+C
\end{equation}
Exponentiating induces the kernel of the multivariate Gaussian density, leading to the VI update of $ q(\bm\beta)=\mathcal{N}\left(\bm\mu_{q(\bm\beta)},\bm\Sigma_{q(\bm\beta)}\right) $ where
\begin{equation}
\bm\Sigma_{q(\bm\beta)}=\textbf{A}^{-1}=\left(\mathbb{E}_{q(\sigma^2)}\left[1/\sigma^2\right]\textbf{X}^T\textbf{X}+\bm\Sigma^{-1}\right)^{-1}
\end{equation}
\begin{align}
\bm\mu_{q(\bm\beta)}=-\dfrac{1}{2}\textbf{A}^{-1}\textbf{B}=\bm\Sigma_{q(\bm\beta)}\left(\mathbb{E}_{q(\sigma^2)}\left[1/\sigma^2\right]\textbf{X}^T\left(\textbf{y}-\mathbb{E}_{q(\textbf{h})}\left[\textbf{h}\right]\right)+\bm\Sigma^{-1}\bm\mu\right)
\end{align}
where $ \mathbb{E}_{q(\textbf{h})}\left[\textbf{h}\right]=\bm\mu_{q(\textbf{h})} $ will be derived in the next section. It is clear that $ \mathbb{E}_{q(\bm\beta)}\left[\bm\beta\right]=\bm\mu_{q(\bm\beta)} $.

\subsection*{VI Update for $ \textbf{h} $}
Deriving the MFVI updates for $ \textbf{h} $ proceeds similarly by first defining the conditional posterior
\begin{equation}
\pi\left(\textbf{h}|\bm\beta,\sigma^2,\tau\right)\propto\exp\left[-\dfrac{1}{2}\left(\textbf{y}-\textbf{h}-\textbf{X}\bm\beta\right)^T\left(\textbf{y}-\textbf{h}-\textbf{X}\bm\beta\right)/\sigma^2\right]\times\exp\left[-\dfrac{1}{2}\textbf{h}^T\textbf{K}^{-1}\textbf{h}/\tau\right]
\end{equation}
The expectation of the log conditional posterior is therefore
\begin{align}
\mathbb{E}_{-q(\textbf{h})}\left[\ln\pi\left(\textbf{h}|\bm\beta,\sigma^2,\tau\right)\right]=-\dfrac{1}{2}\mathbb{E}_{q(\bm\beta)}\left[\left(\textbf{y}-\textbf{h}-\textbf{X}\bm\beta\right)^T\left(\textbf{y}-\textbf{h}-\textbf{X}\bm\beta\right)\right]\times\mathbb{E}_{q(\sigma^2)}\left[1/\sigma^2\right]
\\\nonumber-\dfrac{1}{2}\textbf{h}^T\textbf{K}^{-1}\textbf{h}\times\mathbb{E}_{q(\tau)}\left[1/\tau\right]+C
\\\nonumber=-\dfrac{1}{2}\left[\text{tr}\left(\mathbb{V}_{q(\bm\beta)}\left[\textbf{X}\bm\beta\right]\right)+\left(\textbf{y}-\textbf{h}-\textbf{X}\times\mathbb{E}_{q(\bm\beta)}\left[\bm\beta\right]\right)^T\left(\textbf{y}-\textbf{h}-\textbf{X}\times\mathbb{E}_{q(\bm\beta)}\left[\bm\beta\right]\right)\right]\times\mathbb{E}_{q(\sigma^2)}\left[1/\sigma^2\right]
\\\nonumber-\dfrac{1}{2}\textbf{h}^T\textbf{K}^{-1}\textbf{h}\times\mathbb{E}_{q(\tau)}\left[1/\tau\right]+C
\\\nonumber=-\dfrac{1}{2}\left(\textbf{y}-\textbf{h}-\textbf{X}\bm\mu_{q(\bm\beta)}\right)^T\left(\textbf{y}-\textbf{h}-\textbf{X}\bm\mu_{q(\bm\beta)}\right)\times\mathbb{E}_{q(\sigma^2)}\left[1/\sigma^2\right]-\dfrac{1}{2}\textbf{h}^T\textbf{K}^{-1}\textbf{h}\times\mathbb{E}_{q(\tau)}\left[1/\tau\right]+C
\end{align}

Letting $ \bm\epsilon=\textbf{y}-\textbf{X}\bm\mu_{q(\bm\beta)} $, the quadratic can be simplified as
\begin{align}
\mathbb{E}_{-q(\textbf{h})}\left[\ln\pi\left(\textbf{h}|\bm\beta,\sigma^2,\tau\right)\right]=-\dfrac{1}{2}\left[\textbf{h}^T\left(\bm\Sigma_0^{-1}+\mathbb{E}_{q(\tau)}\left[1/\tau\right]\textbf{K}^{-1}\right)\textbf{h}-2\left(\bm\epsilon^T\bm\Sigma_0^{-1}\right)\textbf{h}+C\right]
\end{align}
Setting $ \textbf{A}=\bm\Sigma_0^{-1}+\mathbb{E}_{q(\tau)}\left[1/\tau\right]\textbf{K}^{-1} $ and $ \textbf{B}=-2\left(\bm\epsilon^T\bm\Sigma_0^{-1}\right)^T $ allows the square to be completed, giving rise to the Gaussian update
\begin{equation}
\bm\Sigma_{q(\textbf{h})}=\left(\textbf{I}\times\mathbb{E}_{q(\sigma^2)}\left[1/\sigma^2\right]+\mathbb{E}_{q(\tau)}\left[1/\tau\right]\textbf{K}^{-1}\right)^{-1}
\end{equation}
\begin{equation}
\bm\mu_{q(\textbf{h})}=\bm\Sigma_{q(\textbf{h})}\left(\textbf{y}-\textbf{X}\bm\mu_{q(\bm\beta)}\right)\times\mathbb{E}_{q(\sigma^2)}\left[1/\sigma^2\right]
\end{equation}
Thus $ \mathbb{E}_{q(\textbf{h})}\left[\textbf{h}\right]=\bm\mu_{q(\textbf{h})} $.

\subsection*{VI Update for $ \bm\sigma^2 $}
The MFVI updates for $ \sigma^2 $ are much simpler to derive since there is no need to complete the square. Once again the conditional posterior is
\begin{align}
\pi\left(\sigma^2|\textbf{h},\bm\beta\right)\propto\det\left[\sigma^2\textbf{I}_n\right]^{-1/2}\exp\left[-\dfrac{1}{2}\left(\textbf{y}-\textbf{h}-\textbf{X}\bm\beta\right)^T\left(\textbf{y}-\textbf{h}-\textbf{X}\bm\beta\right)/\sigma^2\right]
\\\nonumber\times\left(\sigma^2\right)^{-1-\nu_\sigma/2}\exp\left[-\dfrac{\nu_\sigma\sigma_0^2}{2\sigma^2}\right]
\\\nonumber=\left(\sigma^2\right)^{-\left(1+\dfrac{n+\nu_\sigma}{2}\right)}\exp\left[-\dfrac{1}{2}\left(\textbf{y}-\textbf{h}-\textbf{X}\bm\beta\right)^T\left(\textbf{y}-\textbf{h}-\textbf{X}\bm\beta\right)/\sigma^2-\dfrac{1}{2}\nu_\sigma\sigma_0^2/\sigma^2\right]
\end{align}
Computing the log-expectation is relatively straightforward
\begin{align}
\mathbb{E}_{-q(\sigma^2)}\left[\ln\pi\left(\sigma^2|\textbf{h},\bm\beta\right)\right]=-\left(1+\dfrac{n+\nu_\sigma}{2}\right)\ln\sigma^2
\\\nonumber-\dfrac{1}{2}\left[\text{tr}\left(\bm\Sigma_{q(\textbf{h})}+\textbf{X}\bm\Sigma_{q(\bm\beta)}\textbf{X}^T\right)+\left(\textbf{y}-\bm\mu_{q(\textbf{h})}-\textbf{X}\bm\mu_{q(\bm\beta)}\right)^T\left(\textbf{y}-\bm\mu_{q(\textbf{h})}-\textbf{X}\bm\mu_{q(\bm\beta)}\right)\right]/\sigma^2
\\\nonumber-\dfrac{1}{2}\nu_\sigma\sigma_0^2/\sigma^2+C
\end{align}
Letting
\begin{equation}
D=\text{tr}\left(\bm\Sigma_{q(\textbf{h})}+\textbf{X}\bm\Sigma_{q(\bm\beta)}\textbf{X}^T\right)+\left(\textbf{y}-\bm\mu_{q(\textbf{h})}-\textbf{X}\bm\mu_{q(\bm\beta)}\right)^T\left(\textbf{y}-\bm\mu_{q(\textbf{h})}-\textbf{X}\bm\mu_{q(\bm\beta)}\right)
\end{equation}
allows for the simplification
\begin{equation}
\mathbb{E}_{-q(\sigma^2)}\left[\ln\pi\left(\sigma^2|\textbf{h},\bm\beta\right)\right]=-\left(1+\dfrac{n+\nu_\sigma}{2}\right)\ln\sigma^2-\dfrac{D+\nu_\sigma\sigma_0^2}{2\sigma^2}+C
\end{equation}
which is the log-kernel of the scaled-inverse chi-squared distribution. The MFVI updates are therefore
\begin{equation}
\nu_{q(\sigma^2)}=n+\nu_\sigma
\end{equation}
\begin{equation}
\sigma^2_{0,q(\sigma^2)}=\dfrac{D+\nu_\sigma\sigma_0^2}{\nu_{q(\sigma^2)}}
\end{equation}
This makes it easy to calculate
\begin{equation}
\mathbb{E}_{q(\sigma^2)}\left[1/\sigma^2\right]=1/\sigma_{0,q(\sigma^2)}^2
\end{equation}

\subsection*{VI Update for $ \bm\tau $}
As usual, the conditional posterior is calculated
\begin{align}
\pi\left(\tau|\textbf{h}\right)\propto\det\left[\tau\textbf{K}\right]^{-1/2}\exp\left[-\dfrac{1}{2}\textbf{h}^T\textbf{K}^{-1}\textbf{h}/\tau\right]\times\tau^{-1-\nu_\tau/2}\exp\left[-\dfrac{\nu_\tau\tau_0}{2\tau}\right]
\\\nonumber\propto\tau^{-\left(1+\dfrac{n+\nu_\tau}{2}\right)}\exp\left[-\dfrac{1}{2}\textbf{h}^T\textbf{K}^{-1}\textbf{h}/\tau-\dfrac{1}{2}\nu_\tau\tau_0/\tau\right]
\end{align}
The conditional log-posterior is
\begin{align}
\ln\pi\left(\tau|\textbf{h}\right)=-\left(1+\dfrac{n+\nu_\tau}{2}\right)\ln\tau-\dfrac{1}{2}\textbf{h}^T\textbf{K}^{-1}\textbf{h}/\tau-\dfrac{1}{2}\nu_\tau\tau_0/\tau+C
\end{align}

Taking expectations, the log expectation can be written as
\begin{align}
\mathbb{E}_{-q(\tau)}\left[\ln\pi\left(\tau|\textbf{h}\right)\right]=-\left(1+\dfrac{n+\nu_\tau}{2}\right)\ln\tau-\dfrac{1}{2}\mathbb{E}_{q(\textbf{h})}\left[\textbf{h}^T\textbf{K}^{-1}\textbf{h}\right]/\tau-\dfrac{1}{2}\nu_\tau\tau_0/\tau+C
\end{align}
This can be simplified by substituting
\begin{equation}
D=\mathbb{E}_{q(\textbf{h})}\left[\textbf{h}^T\textbf{K}^{-1}\textbf{h}\right]
=\text{tr}\left(\textbf{K}^{-1}\bm{\Sigma}_{q(\textbf{h})}\right)+\bm\mu_{q(\textbf{h})}^T\textbf{K}^{-1}\bm\mu_{q(\textbf{h})}
\end{equation}
and the log-expectation is finally
\begin{align}
\mathbb{E}_{-q(\tau)}\left[\ln\pi\left(\tau|\textbf{h},\rho\right)\right]=-\left(1+\dfrac{n+\nu_\tau}{2}\right)\ln\tau-\dfrac{1}{2}D/\tau-\dfrac{1}{2}\nu_\tau\tau_0/\tau+C
\end{align}
which is again the log-kernel of the scaled-inverse chi-squared distribution with MFVI updates
\begin{equation}
\nu_{q(\tau)}=n+\nu_\tau
\end{equation}
\begin{equation}
\tau_{0,q(\tau)}=\dfrac{D+\nu_\tau\tau_0}{\nu_{q(\tau)}}
\end{equation}
allowing for the calculation of the expectation
\begin{equation}
\mathbb{E}_{q(\tau)}\left[1/\tau\right]=1/\tau_{0,q(\tau)}
\end{equation}

\subsection*{Assessing Convergence}
Convergence of MFVI is traditionally assessed by monitoring the KL divergence between the posterior $ p(\bm\theta|\textbf{y}) $ and the approximation $ q(\bm\theta) $. In this case
\begin{equation}
\mathcal{L}=\text{KL}(q||p)=-\mathbb{E}_q\left[\ln\dfrac{p(\bm\theta|\textbf{y})}{q(\bm\theta)}\right]=\mathbb{E}_q\left[\ln q(\bm\theta)\right]-\mathbb{E}_q\left[p(\bm\theta|\textbf{y})\right]
\end{equation}
The first term is relatively easy to calculate since $ q $ is a fully factored distribution over each parameter of interest. Since $ \mathbb{E}_q\left[q(\bm\theta)\right]=\mathbb{E}_{q(\bm\beta)}\left[q(\bm\beta)\right]+\mathbb{E}_{q(\textbf{h})}\left[q(\textbf{h})\right]+\mathbb{E}_{q(\sigma^2)}\left[q(\sigma^2)\right]+\mathbb{E}_{q(\tau)}\left[q(\tau)\right] $ the first term is just the sum of the negative entropies which are known quantities for normal and scaled-inverse chi-squared distributions
\begin{equation}
\mathbb{E}_{q(\bm\beta)}\left[q(\bm\beta)\right]=-\dfrac{1}{2}\ln\det\left[\bm\Sigma_{q(\bm\beta)}\right]+C
\end{equation}
\begin{equation}
\mathbb{E}_{q(\textbf{h})}\left[q(\textbf{h})\right]=-\dfrac{1}{2}\ln\det\left[\bm\Sigma_{q(\textbf{h})}\right]+C
\end{equation}
The entropy for the scaled-inverse chi-squared distributions are a little more complicated, however only the scale parameters $ \sigma_{0,q(\sigma^2)}^2 $ and $ \tau_{0,q(\tau)} $ are updated with each MFVI iteration, so for the purposes of tracking convergence all terms involving the degrees of freedom parameters $ \nu_{q(\sigma^2)} $ and $ \nu_{q(\tau)} $ can be included in the additive constant term
\begin{equation}
\mathbb{E}_{q(\sigma^2)}\left[q(\sigma^2)\right]=-\ln\sigma_{0,q(\sigma^2)}^2+C
\end{equation}
\begin{equation}
\mathbb{E}_{q(\tau)}\left[q(\tau)\right]=-\ln\tau_{0,q(\tau)}+C
\end{equation}

To derive the $ \mathbb{E}_q\left[p(\bm\theta|\textbf{y})\right] $ term, it is first helpful to write out the full posterior
\begin{align}
\pi\left(\bm\beta,\textbf{h},\sigma^2,\tau\right)\propto\det\left[\textbf{I}_n\sigma^2\right]^{-1/2}\exp\left[-\dfrac{1}{2}\left(\textbf{y}-\textbf{h}-\textbf{X}\bm\beta\right)^T\left(\textbf{y}-\textbf{h}-\textbf{X}\bm\beta\right)/\sigma^2\right]
\\\nonumber\times\exp\left[-\dfrac{1}{2}\left(\bm\beta-\bm\mu\right)^T\bm\Sigma^{-1}\left(\bm\beta-\bm\mu\right)\right]\det\left[\tau\textbf{K}\right]^{-1/2}\exp\left[-\dfrac{1}{2}\textbf{h}^T\textbf{K}^{-1}\textbf{h}/\tau\right]
\\\nonumber\times\left(\sigma^2\right)^{-\left(1+\nu_\sigma/2\right)}\exp\left[-\dfrac{\nu_\sigma\sigma_0^2}{2\sigma^2}\right]\tau^{-\left(1+\nu_\tau/2\right)}\exp\left[-\dfrac{\nu_\tau\tau_0}{2\tau}\right]
\\\nonumber\propto \left(\sigma^2\right)^{-\left(1+\dfrac{\nu_\sigma+n}{2}\right)}\tau^{-\left(1+\dfrac{\nu_\tau+n}{2}\right)}\exp\Big[-\dfrac{1}{2}\left(\textbf{y}-\textbf{h}-\textbf{X}\bm\beta\right)^T\left(\textbf{y}-\textbf{h}-\textbf{X}\bm\beta\right)/\sigma^2
\\\nonumber-\dfrac{1}{2}\left(\bm\beta-\bm\mu\right)^T\bm\Sigma^{-1}\left(\bm\beta-\bm\mu\right)-\dfrac{1}{2}\textbf{h}^T\textbf{K}^{-1}\textbf{h}/\tau-\dfrac{\nu_\sigma\sigma_0^2}{2\sigma^2}-\dfrac{\nu_\tau\tau_0}{2\tau}\Big]
\end{align}

When taking the log-expectation over $ q $, the difficult expectation will be of $ \ln\sigma^2 $ and $ \ln\tau $. For a random variable $ z $ with a scaled-inverse chi-squared distribution with degrees of freedom $ \nu $ and scale parameter $ s^2 $, the expectation is $ \mathbb{E}\left[\ln z\right]=\ln\left(\nu s^2/2\right)-\psi\left(\nu/2\right) $. With the $ q $-densities, the degrees of freedom parameters remain the same for each MFVI iteration, so for the purposes of monitoring convergence, the expectations can be calculated as
\begin{equation}
\mathbb{E}_{q(\sigma^2)}\left[\ln\sigma^2\right]=\ln\sigma_{0,q(\sigma^2)}^2+C
\end{equation}
\begin{equation}
\mathbb{E}_{q(\tau)}\left[\ln\tau\right]=\ln\tau_{0,q(\tau)}+C
\end{equation}
Applying all of these facts yields the log-expectation
\begin{align}
\mathbb{E}_q\left[\pi(\bm\theta|\textbf{y})\right]=-\left(1+\dfrac{\nu_\sigma+n}{2}\right)\ln\sigma_{0,q(\sigma^2)}^2-\left(1+\dfrac{\nu_\tau+n}{2}\right)\ln\tau_{0,q(\tau)}
\\\nonumber-\dfrac{1}{2}\Big[
\text{tr}\left(\bm\Sigma_{q(\textbf{h})}+\textbf{X}\bm\Sigma_{q(\bm\beta)}\textbf{X}^T\right)/\sigma_{0,q(\sigma^2)}^2+\left(\textbf{y}-\bm\mu_{q(\textbf{h})}-\textbf{X}\bm\mu_{q(\bm\beta)}\right)^T\left(\textbf{y}-\bm\mu_{q(\textbf{h})}-\textbf{X}\bm\mu_{q(\bm\beta)}\right)/\sigma_{0,q(\sigma^2)}^2
\\\nonumber+
\text{tr}\left(\bm\Sigma^{-1}\bm\Sigma_{q(\bm\beta)}\right)+\left(\bm\mu_{q(\bm\beta)}-\bm\mu\right)\bm\Sigma^{-1}\left(\bm\mu_{q(\bm\beta)}-\bm\mu\right)+\text{tr}\left(\textbf{K}^{-1}\bm\Sigma_{q(\textbf{h})}\right)/\tau_{0,q(\tau)}+\bm\mu_{q(\textbf{h})}^T\textbf{K}^{-1}\bm\mu_{q(\textbf{h})}/\tau_{0,q(\tau)}
\\\nonumber+\dfrac{\nu_\sigma\sigma_0^2}{\sigma_{0,q(\sigma^2)}^2}+\dfrac{\nu_\tau\tau_0}{\tau_{0,q(\tau)}}\Big]+C
\end{align}
which completes the calculations necessary to track convergence of MFVI via $ \mathcal{L} $.

\newpage
\section*{Appendix: VI Updates with Flat Priors}
\subsection*{VI Update for $ \bm\beta $}
Beginning with the MFVI updates for $ \bm\beta $, the conditional posterior is
\begin{equation}
\pi\left(\bm\beta|\textbf{h},\sigma^2\right)\propto\exp\left[-\dfrac{1}{2}\left(\textbf{y}-\textbf{h}-\textbf{X}\bm\beta\right)^T\left(\textbf{y}-\textbf{h}-\textbf{X}\bm\beta\right)/\sigma^2\right]
\end{equation}
The expectation of the log conditional posterior is therefore
\begin{align}
\mathbb{E}_{-q(\bm\beta)}\left[\ln\pi\left(\bm\beta|\textbf{h},\sigma^2\right)\right]=-\dfrac{1}{2}\mathbb{E}_{q(\textbf{h})}\left[\left(\textbf{y}-\textbf{h}-\textbf{X}\bm\beta\right)^T\left(\textbf{y}-\textbf{h}-\textbf{X}\bm\beta\right)\right]\times\mathbb{E}_{q(\sigma^2)}\left[1/\sigma^2\right]+C
\\\nonumber=-\dfrac{1}{2}\left[\text{tr}\left(\mathbb{V}_{q(\textbf{h})}\left[\textbf{h}\right]\right)+\left(\textbf{y}-\mathbb{E}_{q(\textbf{h})}\left[\textbf{h}\right]-\textbf{X}\bm\beta\right)^T\left(\textbf{y}-\mathbb{E}_{q(\textbf{h})}\left[\textbf{h}\right]-\textbf{X}\bm\beta\right)\right]\times\mathbb{E}_{q(\sigma^2)}\left[1/\sigma^2\right]+C
\\\nonumber=-\dfrac{1}{2}\left(\textbf{y}-\bm\mu_{q(\textbf{h})}-\textbf{X}\bm\beta\right)^T\left(\textbf{y}-\bm\mu_{q(\textbf{h})}-\textbf{X}\bm\beta\right)/\sigma_{0,q(\sigma^2)}^2+C
\end{align}

Let $ \textbf{d}=\textbf{y}-\bm\mu_{q(\textbf{h})} $ and $ \Sigma_0^{-1}=\textbf{I}_n/\sigma_{0,q(\sigma^2)}^2 $, the entire expression can be simplified into a single quadratic form
\begin{equation}
\mathbb{E}_{-q(\bm\beta)}\left[\ln\pi\left(\bm\beta|\textbf{h},\sigma^2\right)\right]=-\dfrac{1}{2}\left[\bm\beta^T\left(\textbf{X}^T\bm\Sigma_0^{-1}\textbf{X}\right)\bm\beta-2\left(\textbf{d}^T\bm\Sigma_0^{-1}\textbf{X}\right)\bm\beta+C\right]
\end{equation}

Letting $ \textbf{A}=\textbf{X}^T\bm\Sigma_0^{-1}\textbf{X}=\textbf{X}^T\textbf{X}/\sigma_{0,q(\sigma^2)}^2 $ and $ \textbf{B}=-2\textbf{X}^T\bm\Sigma_0^{-1}\textbf{d} =-2\textbf{X}^T\textbf{d}/\sigma_{0,q(\sigma^2)}^2$, the square can be completed, yielding the updates
\begin{equation}
\bm\Sigma_{q(\bm\beta)}=\textbf{A}^{-1}=\left(\textbf{X}^T\textbf{X}\right)^{-1}\sigma_{0,q(\sigma^2)}^2
\end{equation}
\begin{align}
\bm\mu_{q(\bm\beta)}=-\dfrac{1}{2}\textbf{A}^{-1}\textbf{B}=\bm\Sigma_{q(\bm\beta)}\textbf{X}^T\left(\textbf{y}-\bm\mu_{q(\textbf{h})}\right)/\sigma_{0,q(\sigma^2)}^2
\end{align}

\subsection*{VI Update for $ \textbf{h} $}
In examining the conditional posterior for $ \textbf{h} $, it is clear that it takes the same form as in the case with informative priors. The VI updates are therefore the same, i.e.
\begin{equation}
\bm\Sigma_{q(\textbf{h})}=\left(\textbf{I}/\sigma_{0,q(\sigma^2)}^2+\textbf{K}^{-1}/\tau_{0,q(\tau)}\right)^{-1}
\end{equation}
\begin{equation}
\bm\mu_{q(\textbf{h})}=\bm\Sigma_{q(\textbf{h})}\left(\textbf{y}-\textbf{X}\bm\mu_{q(\bm\beta)}\right)/\sigma_{0,q(\sigma^2)}^2
\end{equation}

\subsection*{VI Update for $ \bm\sigma^2 $}
The MFVI updates for $ \sigma^2 $ are much simpler to derive since there is no need to complete the square. Once again the conditional posterior is
\begin{align}
\pi\left(\sigma^2|\textbf{h},\bm\beta\right)\propto\det\left[\sigma^2\textbf{I}_n\right]^{-1/2}\exp\left[-\dfrac{1}{2}\left(\textbf{y}-\textbf{h}-\textbf{X}\bm\beta\right)^T\left(\textbf{y}-\textbf{h}-\textbf{X}\bm\beta\right)/\sigma^2\right]
\\\nonumber\propto\left(\sigma^2\right)^{-n/2}\exp\left[-\dfrac{1}{2}\left(\textbf{y}-\textbf{h}-\textbf{X}\bm\beta\right)^T\left(\textbf{y}-\textbf{h}-\textbf{X}\bm\beta\right)/\sigma^2\right]
\end{align}
Now note that this is again the kernel of the scaled-inverse chi-squared distribution with
\begin{equation}
\left(\sigma^2\right)^{-n/2}=\left(\sigma^2\right)^{-\left(1+\nu_{q(\sigma^2)}/2\right)}
\end{equation}
which means that the degrees of freedom is $ \nu_{q(\sigma^2)}=n-2 $.

Computing the log-expectation is relatively straightforward
\begin{align}
\mathbb{E}_{-q(\sigma^2)}\left[\ln\pi\left(\sigma^2|\textbf{h},\bm\beta\right)\right]=-\dfrac{n}{2}\ln\sigma^2
\\\nonumber-\dfrac{1}{2}\left[\text{tr}\left(\bm\Sigma_{q(\textbf{h})}+\textbf{X}\bm\Sigma_{q(\bm\beta)}\textbf{X}^T\right)+\left(\textbf{y}-\bm\mu_{q(\textbf{h})}-\textbf{X}\bm\mu_{q(\bm\beta)}\right)^T\left(\textbf{y}-\bm\mu_{q(\textbf{h})}-\textbf{X}\bm\mu_{q(\bm\beta)}\right)\right]/\sigma^2+C
\end{align}
Letting
\begin{equation}
D=\text{tr}\left(\bm\Sigma_{q(\textbf{h})}+\textbf{X}\bm\Sigma_{q(\bm\beta)}\textbf{X}^T\right)+\left(\textbf{y}-\bm\mu_{q(\textbf{h})}-\textbf{X}\bm\mu_{q(\bm\beta)}\right)^T\left(\textbf{y}-\bm\mu_{q(\textbf{h})}-\textbf{X}\bm\mu_{q(\bm\beta)}\right)
\end{equation}
allows for the simplification
\begin{equation}
\mathbb{E}_{-q(\sigma^2)}\left[\ln\pi\left(\sigma^2|\textbf{h},\bm\beta\right)\right]=-\dfrac{n}{2}\ln\sigma^2-\dfrac{D}{2\sigma^2}+C
\end{equation}

For this to be the log-kernel of scaled-inverse chi-squared distribution we require that $ D=\nu_{q(\sigma^2)}\sigma_{0,q(\sigma^2)} $ that implies $ \sigma_{0,q(\sigma^2)}=D/\nu_{q(\sigma^2)} $ thus completing the VI update.

\subsection*{VI Update for $ \bm\tau $}
The conditional posterior is
\begin{equation}
\pi\left(\tau|\textbf{h}\right)\propto\det\left[\tau\textbf{K}\right]^{-1/2}\exp\left[-\dfrac{1}{2}\textbf{h}^T\textbf{K}^{-1}\textbf{h}/\tau\right]\propto \tau^{-n/2}\exp\left[-\dfrac{1}{2}\textbf{h}^T\textbf{K}^{-1}\textbf{h}/\tau\right]
\end{equation}
Recognizing again that this is the scaled-inverse chi-squared kernel, we see from the first term that $ -n/2=-(1+\nu_{q(\tau)}/2) $, i.e. $ \nu_{q(\tau)}=n-2 $.

The conditional log-posterior is
\begin{equation}
\ln\pi\left(\tau|\textbf{h}\right)=-\dfrac{n}{2}\ln\tau-\dfrac{1}{2}\textbf{h}^T\textbf{K}^{-1}\textbf{h}/\tau+C
\end{equation}

Taking expectations, the log expectation can be written as
\begin{align}
\mathbb{E}_{-q(\tau)}\left[\ln\pi\left(\tau|\textbf{h}\right)\right]=-\dfrac{n}{2}\ln\tau-\dfrac{1}{2}\mathbb{E}_{q(\textbf{h})}\left[\textbf{h}^T\textbf{K}^{-1}\textbf{h}\right]/\tau+C
\end{align}
This can be simplified by substituting
\begin{equation}
D=\mathbb{E}_{q(\textbf{h})}\left[\textbf{h}^T\textbf{K}^{-1}\textbf{h}\right]
=\text{tr}\left(\textbf{K}^{-1}\bm{\Sigma}_{q(\textbf{h})}\right)+\bm\mu_{q(\textbf{h})}^T\textbf{K}^{-1}\bm\mu_{q(\textbf{h})}
\end{equation}
and the log-expectation is finally
\begin{align}
\mathbb{E}_{-q(\tau)}\left[\ln\pi\left(\tau|\textbf{h},\rho\right)\right]=-\dfrac{n}{2}\ln\tau-\dfrac{1}{2}D/\tau+C
\end{align}
which is again the log-kernel of the scaled-inverse chi-squared distribution. Since we require we require that $ D=\nu_{q(\tau)}\tau_{0,q(\tau)} $ that implies $ \tau_{0,q(\tau)}=D/\nu_{q(\tau)} $.

\subsection*{Assessing Convergence}
Note that the posterior in the flat prior case takes the same parametric form as in the informative prior case, i.e. Gaussian posteriors for $ \textbf{h} $ and $ \bm\beta $ and scaled-inverse chi-squared posteriors for $ \sigma^2 $ and $ \tau $. Thus $ \mathbb{E}_q\left[\ln q(\bm\theta)\right] $ is of the same form so only $ \mathbb{E}_q\left[p(\bm\theta|\textbf{y})\right] $ must be calculated.

The full posterior is
\begin{align}
\pi\left(\bm\beta,\textbf{h},\sigma^2,\tau\right)\propto\left(\sigma^2\right)^{-n/2}\exp\left[-\dfrac{1}{2}\left(\textbf{y}-\textbf{h}-\textbf{X}\bm\beta\right)^T\left(\textbf{y}-\textbf{h}-\textbf{X}\bm\beta\right)/\sigma^2\right]
\\\nonumber\tau^{-n/2}\exp\left[-\dfrac{1}{2}\textbf{h}^T\textbf{K}^{-1}\textbf{h}/\tau\right]
\end{align}
and applying the same properties as in the informative prior case, we obtain the log-expectation
\begin{align}
\mathbb{E}_q\left[\pi(\bm\theta|\textbf{y})\right]=-\dfrac{n}{2}\ln\sigma_{0,q(\sigma^2)}^2-\dfrac{n}{2}\ln\tau_{0,q(\tau)}
\\\nonumber-\dfrac{1}{2}\Big[
\text{tr}\left(\bm\Sigma_{q(\textbf{h})}+\textbf{X}\bm\Sigma_{q(\bm\beta)}\textbf{X}^T\right)/\sigma_{0,q(\sigma^2)}^2+\left(\textbf{y}-\bm\mu_{q(\textbf{h})}-\textbf{X}\bm\mu_{q(\bm\beta)}\right)^T\left(\textbf{y}-\bm\mu_{q(\textbf{h})}-\textbf{X}\bm\mu_{q(\bm\beta)}\right)/\sigma_{0,q(\sigma^2)}^2
\\\nonumber+
\text{tr}\left(\textbf{K}^{-1}\bm\Sigma_{q(\textbf{h})}\right)\tau_{0,q(\tau)}+\bm\mu_{q(\textbf{h})}^T\textbf{K}^{-1}\bm\mu_{q(\textbf{h})}/\tau_{0,q(\tau)}\Big]
\end{align}
which completes the calculations necessary to track convergence of MFVI via $ \mathcal{L} $.

\end{document}